\documentclass[twocolumn,showpacs,amsmath,amssymb,prl]{revtex4}

\usepackage{graphicx}
\usepackage{dcolumn}
\usepackage{bm}
\usepackage{epsfig}

\begin{document}

\textbf{Comment on "Potential Energy Landscape for Hot Electrons in Periodically Nanostructured Graphene"}\\

In a recent letter \cite{bor10} the unoccupied electronic states of single layers of graphene ($g$) on ruthenium are investigated. 
It is shown that elevated graphene areas with a diameter of about 2 nm ($H$-areas or "hills") look like quantum dots \cite{bor10,zha10}. 
Regions where graphene binds to the substrate are the $L$-areas or "valleys", and separate the dots.

\begin{figure}[b]
\includegraphics[width=8 cm]{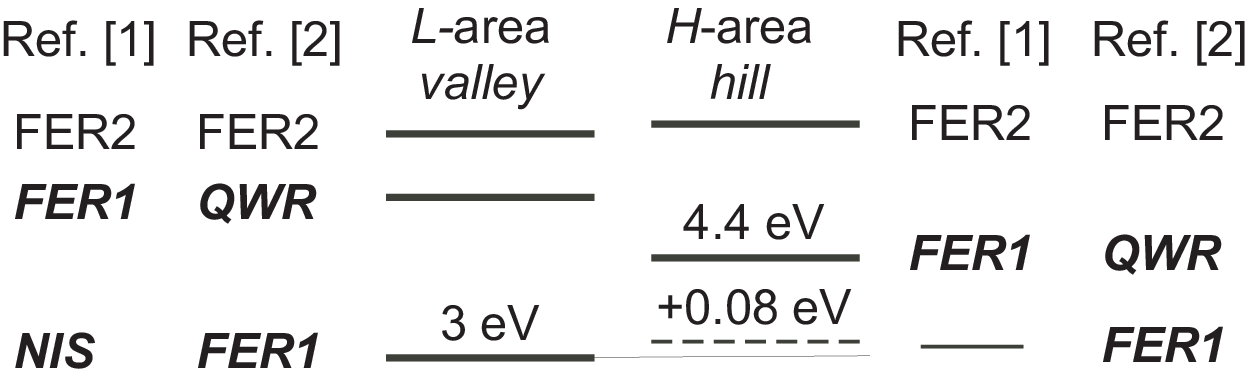}
\caption{Peak assignments in the $L$- and $H$-areas of $g$/Ru(0001). The first two Field Emission Resonances (FER1 and FER2), the New Interfacial State (NIS) of Ref.\cite{bor10} and the Quantum Well Resonance (QWR) from Ref.\cite{zha10} are marked. 
}
\label{F1}
\end{figure}

\begin{figure}
\includegraphics[width=8.5cm]{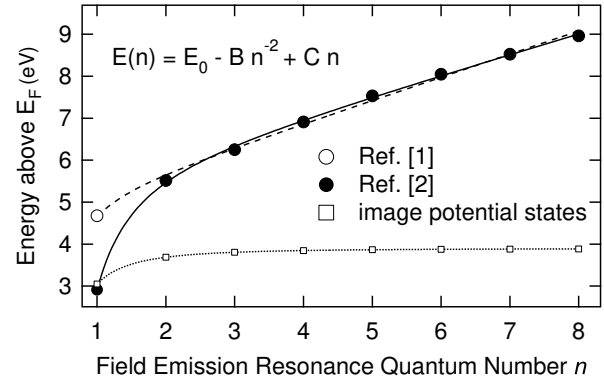}
\caption{Field Emission Resonance energies (FER1 to FER8) on the $L$-areas in $g$/Ru(0001) (Energy values from \cite{bor10}). Both assignments Ref.\cite{bor10} and Ref.\cite{zha10} may be fitted to an empirical trend $E(n)=E_0-Bn^{-2}+Cn$. For comparison an image potential state series for $E_0$=3.9 eV, $B$=0.85 eV and $C$=0 is shown.}
\label{F2}
\end{figure}

Here we comment on the two interpretations \cite{bor10} and \cite{zha10}, which deviate in four points (see Figure \ref{F1}) and outline the corresponding consequences.
First we want to discuss the assignment of the first field emission resonance (FER1) that originates from the first image potential state 0.85 eV below the vacuum level \cite{bin85,str86}.
The 3 eV peak in the $L$-area is assigned in Ref.\cite{zha10} to the first field emission resonance, with a corresponding state at a significantly higher energy (0.08$\pm 0.03$ eV) in the $H$-area. 
Note that the different energies of FER1 on the hills and in the valley and the discrimination of higher FER's on $H$ and $L$ exclude this peak to be related to cross talk.
Figure \ref{F2} shows however that both assignments of FER$n$ are compatible with an empirical trend $E(n)=E_0-Bn^{-2}+Cn$ of the FER energies. 
Both show smaller $E_0$'s on the $L$-areas and thus indicate a larger local work function on the $H$-areas \cite{bru09}. 
The obtained $B$ values give a hint on which peak identification holds: While \cite{bor10} results in $B$=0.6 eV, interpretation \cite{zha10} results in $B$=2.7 eV.
A $B$ which is larger than the 0.85 eV of the image state, indicates that the $-1/4z$ image potential with a hard wall mirror on the metal side, does not apply, but that there is an anomalous FER energy lowering in the $g$/Ru(0001) system. Ref.\cite{zha10} identifies this lowering to be due to the graphene quantum well, which delocalizes the FER wave functions.
From this the assignment of the other peaks in interpretation \cite{zha10} follows:
The strong resonance on the $H$-areas is a quantum well resonance (QWR) with a counterpart in the $L$-areas at 500 meV \emph{higher} energy, which is due to the thinner well in the $L$-areas.
Furthermore, interpretation \cite{zha10} gives arguments for the peak intensities and has, in the sense of Occams razor, the advantage that \emph{all} FER's follow the trend of the local work functions.

On the other hand, the non-observation of the "new interfacial state" (NIS) on the $H$-areas in Ref.\cite{bor10} leaves a problem. Besides this, the localisation of the electrons in the 4.4 eV peak is "outside" the graphene \cite{bor10} or "inside" \cite{zha10}, which is an essential difference if the state shall be used in the context of a quantum dot array or the dynamics of hot electrons.
Although the concept of the double Rydberg series of free standing graphene \cite{sil09} also imposes new states,
it is not clear whether it is of use for the present situation, because the substrate breaks the symmetry. 
The first principles calculations in \cite{bor10} base on calculations within a (1x1) Ru(0001) cell with strained graphene on top. They do not include the image potential tail nor the lateral localisation (Ref.\cite{zha10} estimates the effect for the hills to be in the order of 0.2 eV). This, and the straining of the graphene could give a substantial change of the energies, with respect to the vacuum and/or Fermi level, and we expect that first principle calculations which include these considerations are consistent with the model and the assignments in  Ref.\cite{zha10}.
\\
\\
H.G. Zhang$^1$, and T. Greber$^{2}$,
$^1$Institute of Physics, Chinese Academy of Sciences, Beijing 100190, China,
$^2$Physik-Institut, University of Z\"urich, 8057 Z\"urich, Switzerland. \today.

\end{document}